# Domain and Phase De-strain – Formation of Ferroelastic Domain Structures

Fei Xue,[1] Yongjun Li,[2] Yijia Gu,[1] Jinxing Zhang,[2] and Long-Qing Chen[1]

[1]Department of Materials Science and Engineering, The Pennsylvania State University, University Park, Pennsylvania 16802, USA

[2]Department of Physics, Beijing Normal University, Beijing, 100875, China

**Abstract:**

Phase decomposition is a well-known process leading to the formation of two-phase mixtures. Here we show that a strain imposed on a ferroelastic crystal promotes the formation of mixed phases and domains, i.e., domain and phase de-strain with local strains determined by a common tangent construction on the free energy versus strain curves. It is demonstrated that a domain structure can be understood using the concepts of domain/phase rule, lever rule, coherent and incoherent de-strain within the de-strain model description, in a complete analogy to phase decomposition. The proposed de-strain model is validated using phase-field simulations and experimental observations of $PbTiO_3$ and $BiFeO_3$ thin films as examples. The de-strain model provides a simple tool to guide and design domain structures of ferroelastic systems.

PACS number: 62.23.St, 77.55.Nv, 77.80.Dj, 81.30.Kf

Phase separation of a homogeneous state into a mixture of two or more phases is the manifestation of a common mode of materials instability. For a phase decomposition process, the presence and local compositions of mixed phases can be illustrated using the geometrical common tangent construction on the free energy versus composition curves (Fig. 1a). It is based on the thermodynamic condition that the chemical potential of each species has to be uniform at equilibrium [1].

The coexistence of domains with different local strains is also a common phenomenon, known to minimize the overall elastic energy in a constrained system. In a ferroelastic system, the transition from high symmetry phase to low symmetry phase gives rise to different variants of the low symmetry phase, which may coexist and form domains structures. Alternatively, the coexistent domains can be two phases with different space groups and distinct physical properties, which may result in enhanced responses under external stimuli due to the transition between the two phases [2,3]. The $a/c$ multi-domains in $PbTiO_3$ (PTO) films belong to the former case [2-5]. PTO shows a first-order transition from a paraelectric cubic phase to a ferroelectric tetragonal phase in single crystals, and different variants of the tetragonal phase form $a/c$ multi-domains in a thin film state [6,7]. On the other hand, the mixed-phase domain structure in compressively strained $BiFeO_3$ (BFO) films is an example of the latter case [8]. BFO bulk materials show space group $R3c$ with polarization and out-of-phase oxygen octahedral tilt along the <111> pseudocubic direction [9]. Under a compressive strain, a tetragonal-like (T-like) phase is stabilized in BFO films, which shows a large $c/a$ ratio of ~1.25 and polarization of ~1.5 C/m$^2$ [8,10]. The T-like phase is not exactly tetragonal, but a monoclinic $M_c$ phase, with polarization along the [$u0v$] direction [11,12]. The mixed-phase regions in BFO films give rise to a huge electromechanical response, due to the transition between the two-phase mixture and the pure T-like phase [8,13]. Generally, the multi-domain structures and related properties are investigated by the analysis of macrostress [14-16] and phase-field simulations [4,17]. However, the macrostress analysis usually assumes that the local strains are fixed, and phase-field simulations rely on extensive numerical computations.

In this Letter, we show that the strain-induced domain structures can be analyzed by the a new concept of phase de-strain, domain/phase separation under a specified strain, similar to the much better known phase decomposition process with a specified overall composition. The common tangent construction, lever rule, and phase rule in the phase decomposition thermodynamics are shown to be equally applicable to the phase de-strain process. Strain-temperature and strain-strain phase diagrams of PTO are calculated based on the common tangent construction and compared to those obtained from phase-field simulations. Domain structures of the mixed rhombohedral-like (R-like) and T-like phases in BFO are obtained using phase-field simulations and explained using the proposed de-strain model [17,18], and the predicted domain morphology and wall orientations demonstrate excellent agreement with piezoresponse force microscopy (PFM) measurements.

We first discuss the thermodynamics of de-strain using a one dimensional (1D) problem as an illustration (Fig. 1c). The system consists of two phase/domain regions $L^\alpha$ and $L^\beta$, with length fraction $f^\alpha$ and $f^\beta$, respectively. Since the total deformation is conserved, the overall strain can be expressed by

$$\varepsilon^0 = f^\alpha \varepsilon^\alpha + f^\beta \varepsilon^\beta, \tag{1}$$

The fractions of the two phases/domains can be determined from the lever rule based on the conservation Eq. (1),

$$f^\alpha = \frac{\varepsilon^\beta - \varepsilon^0}{\varepsilon^\beta - \varepsilon^\alpha}, \tag{2}$$

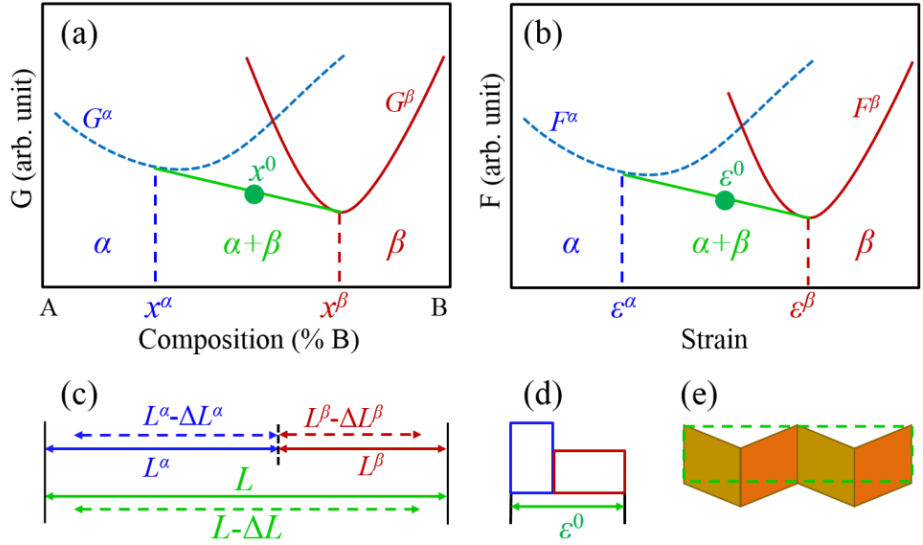

FIG. 1. Schematics of mixed-phases in de-composition and de-strain processes. (a) and (b) Common tangent constructions for (a) phase de-composition and (b) phase de-strain. (c) Overall and local deformation in a 1-D system. (d) and (e) Coexistence of two domains with different (d) normal strains and (e) shear strains.

The free energy of a two-phase mixture is $F^0 = f^\alpha F^\alpha(\varepsilon^\alpha) + (1-f^\alpha)F^\beta(\varepsilon^\beta)$ with $F^\alpha$ and $F^\beta$ the Helmholtz free energy density. Choosing two free variables $\varepsilon^\alpha$ and $f^\alpha$, $\varepsilon^\beta$ can be obtained from Eq. (1), i.e. $\varepsilon^\beta = \frac{\varepsilon^0 - f^\alpha \varepsilon^\alpha}{1 - f^\alpha}$. From the energy minimization conditions $\frac{\partial F^0}{\partial f^\alpha} = 0$ and $\frac{\partial F^0}{\partial \varepsilon^\alpha} = 0$, we have

$$\frac{\partial F^\alpha}{\partial \varepsilon^\alpha} = \frac{\partial F^\beta}{\partial \varepsilon^\beta}, \tag{3}$$

$$\frac{F^\beta(\varepsilon^\beta) - F^\alpha(\varepsilon^\alpha)}{\varepsilon^\beta - \varepsilon^\alpha} = \frac{\partial F^\beta(\varepsilon^\beta)}{\partial \varepsilon^\beta}, \tag{4}$$

Equations (3) and (4) show that the stresses and Gibbs free energy density are uniform at equilibrium, respectively (Gibbs free energy under constrained boundary conditions is given by the Legendre transformation, $G = F - \varepsilon \frac{\partial F}{\partial \varepsilon}$). The chemical potential is related to Gibbs free energy density through $\mu = GV_m$, with $V_m$ the molar volume of the reference state, and thus Eq. (4) also indicates that the chemical potential is uniform, i.e., $\mu^\alpha = \mu^\beta$. From Eqs. (3) and (4), the common tangent of the free energy vs strain curves determines the equilibrium free energy and local strains of the mixed phases/domains (Fig. 1b).

The above discussion for a simple 1D example can be extended to $n$ fixed strain components. A fixed strain can be a normal strain or shear strain, as shown in Figs. 1(d) and 1(e). There are a total of 6 strain/stress components, and a particular mechanical variable component is specified by either a fixed strain or a fixed stress. If there are $n$ fixed strain components, the other 6-$n$ components are specified by stress. With $n$ fixed strains, the common tangent construction results in $n$-D planes in an ($n+1$)-D free energy-strain space, and the lever rule should be modified accordingly [1]. If $p$ types of domains coexist, the equilibrium conditions for stress and chemical potential lead to $n(p-1)$ and $p-1$ constraints, respectively. Thus, the degree of freedom including temperature is

$$d = 1 + np - (n+1)(p-1) = n - p + 2 \tag{5}$$

The domain rule in Eq. (5) is similar to the well-known Gibbs phase rule [19]. The degree of freedom is the number of intensive variables that can be changed independently without disturbing the number of phases in equilibrium. Alternatively, the degree of freedom can be interpreted as the number of local composition or strain components that can be changed independently. It can be employed to predict the maximum possible domains that can co-exist at equilibrium.

We analyze the strain-related phase diagrams of PTO as an example of applying the de-strain model. With polarization $P_i (i=1,2,3)$ as the order parameter, the Helmholtz free energy density of a PTO crystal is given by [18]

$$f_{\text{PTO}} = \alpha_{ij} P_i P_j + \alpha_{ijkl} P_i P_j P_k P_l + \alpha_{ijklmn} P_i P_j P_k P_l P_m P_n + \frac{1}{2} c_{ijkl} (\varepsilon_{ij} - \varepsilon_{ij}^0)(\varepsilon_{kl} - \varepsilon_{kl}^0), \tag{6}$$

where $\alpha_{ij}$, $\alpha_{ijkl}$, and $\alpha_{ijklmn}$ are coefficients of Landau polynomial under stress-free boundary conditions, $c_{ijkl}$ is the elastic stiffness tensor, and $\varepsilon_{ij}$ and $\varepsilon_{ij}^0$ are the total strain and eigen-strain, respectively. The eigen-strain is related to the polarization through $\varepsilon_{ij}^0 = h_{ijkl} P_k P_l$, where $h_{ijkl}$ are coupling coefficients. Note that electrostatic energy is neglected in Eq. (6), which indicates that charge-neutral domain walls and short-circuit electrical boundary conditions are assumed. As a first simplest possible example, 1-D strain $\varepsilon_{22}$ is applied, and other boundary conditions are stress-free,

$$\varepsilon_{22} = \varepsilon_0, \frac{\partial f_{\text{PTO}}}{\partial \varepsilon_{11}} = 0, \frac{\partial f_{\text{PTO}}}{\partial \varepsilon_{12}} = 0, \frac{\partial f_{\text{PTO}}}{\partial \varepsilon_{13}} = 0, \frac{\partial f_{\text{PTO}}}{\partial \varepsilon_{23}} = 0, \frac{\partial f_{\text{PTO}}}{\partial \varepsilon_{33}} = 0, \tag{7}$$

$f_{\text{PTO}}$ can be calculated for different domains, $a_1$ ( $P_1 \neq 0, P_2 = P_3 = 0$ ), $a_2$ ($P_2 \neq 0, P_1 = P_3 = 0$), $c$ ($P_3 \neq 0, P_1 = P_2 = 0$), and $P$ ($P_1 = P_2 = P_3 = 0$). Note that $a_1$ and $c$ domains have the same local $\varepsilon_{22}$, and we choose $c$ domains for the analysis. Using the free energy coefficients from [20], a strain-temperature domain diagram Fig. 2(a) is constructed based on the de-strain model, i.e. by solving Eqs. (3) and (4) under different $\varepsilon_{22}$ and temperature $T$. At temperature $T_0$, the minimal free energies of $c$, $P$, and $a_2$ are equal (middle panel of Fig. 2b), and the three domains may coexist within the strain range between the tangent points of $c$ and $a_2$. Based on the domain rule Eq. (5), the degree of freedom $d = 1+2-3 = 0$, and thus the local strains of the three phases are all fixed. With the increased free energies of $c$ and $a_2$ above $T_0$ (upper panel of Fig. 2b), $P+c$ mixture is favored by a compressive strain, and $P+a_2$ mixture by a tensile strain. Below $T_0$, the free energies of $c$ and $a_2$ are decreased (lower panel of Fig. 2b), and $c+a_2$ mixture is formed under an intermediate strain. Equation (5) is valid for the 2-domain and 1-domain regions in Fig. 2(a), and the strain-temperature domain diagram is similar to a composition-temperature phase diagram of a eutectic binary system [1,19]. Note that the boundaries of $P+c$ and $P+a_2$ domain regions are straight and parallel since only the second order Landau polynomial coefficients are dependent on temperature, and only the electrostrictive coupling is considered between strain and polarization.

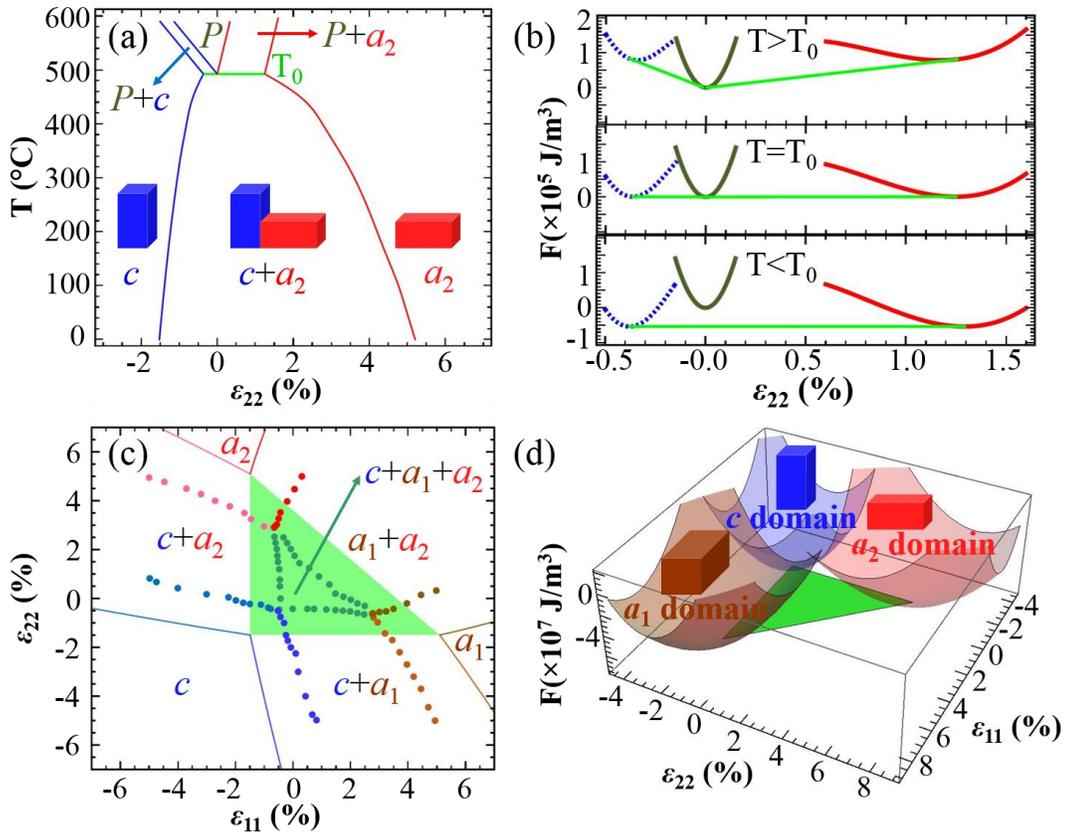
7

FIG. 2. Strain domain diagrams of PbTiO$_3$. (a) Strain-temperature domain diagram. (b) Free energy as a function of strain at different temperatures. (c) Strain-strain domain diagram at room temperature. (d) Energy surfaces of $a_1$, $a_2$, and $c$ domains. The blue triangle denotes the common tangent plane of the three energy surfaces.

As an analogy to the isothermal section of a ternary system [1], strain-strain domain diagrams can be calculated for a specific temperature, under the boundary condition

$$\varepsilon_{11} = \varepsilon_{01}, \varepsilon_{22} = \varepsilon_{02}, \frac{\partial f_{\text{PTO}}}{\partial \varepsilon_{12}} = 0, \frac{\partial f_{\text{PTO}}}{\partial \varepsilon_{13}} = 0, \frac{\partial f_{\text{PTO}}}{\partial \varepsilon_{23}} = 0, \frac{\partial f_{\text{PTO}}}{\partial \varepsilon_{33}} = 0, \tag{8}$$

Fig. 2(d) shows the energy surfaces of $a_1$, $a_2$, and $c$ domains as well as the common tangent plane of the three surfaces at room temperature. Based on the common tangent construction, a strain-strain domain diagram Fig. 2(c) is calculated. Based on Eq. (5), the degrees of freedom in the 3-domain, 2-domain, and 1-domain regions are 0, 1, and 2, respectively. The dots in Fig. 2(c) are data collected from phase-field simulations of the PTO films [5], and the discrepancy is because interfacial energy and additional coherency strain energy are neglected in the common tangent construction [4,21]. Compared with phase-field simulations, the calculation based on common tangent construction requires less computational effort to obtain strain-related domain diagrams, and the determined boundaries between different domain stability regions can be treated as the limiting case when the domain wall density is small, leading to a negligible contribution from interfacial energy.

The second example is the strain induced R/T domains in BFO films. Including the order parameter for oxygen octahedral tilt $\theta_i (i = 1, 2, 3)$, the total free energy density is given by [18,22]

$$f_{\text{BFO}} = \alpha_{ij}P_iP_j + \alpha_{ijkl}P_iP_jP_kP_l + \alpha_{ijklmn}P_iP_jP_kP_lP_mP_n + \alpha_{ijklmnuv}P_iP_jP_kP_lP_mP_nP_uP_v$$
$$+ \beta_{ij}\theta_i\theta_j + \beta_{ijkl}\theta_i\theta_j\theta_k\theta_l + t_{ijkl}P_iP_j\theta_k\theta_l + \frac{1}{2}c_{ijkl}(\varepsilon_{ij} - \varepsilon_{ij}^0)(\varepsilon_{kl} - \varepsilon_{kl}^0),$$
(9)

where $\alpha_{ijklmnuv}$, $\beta_{ij}$, $\beta_{ijkl}$, and $t_{ijkl}$ are coefficients of the Landau polynomial. The eigen-strain is related to the polarization and oxygen octahedral tilt through $\varepsilon_{ij}^0 = \lambda_{ijkl}\theta_k\theta_l + h_{ijkl}P_kP_l$, where $\lambda_{ijkl}$ and $h_{ijkl}$ are coupling coefficients. First-principles calculations show that the T-like monoclinic phase is metastable under stress-free conditions [23]. Thus the polynomial of polarization is expanded to the eighth order, which is necessary to reproduce a stable or metastable monoclinic phase under stress-free conditions [12]. For simplicity, the polynomial of oxygen octahedral tilt and coupling terms are expanded to the lowest order, i.e. the forth order. All the coefficients are listed in supplementary materials, which are fitted based on the experimental measurements and first-principles calculations [23-26].

To confirm whether the Landau polynomial is able to reproduce the T-like phase, the energy landscape is investigated under stress-free conditions, i.e. $\varepsilon_{ij} = \varepsilon_{ij}^0$, eliminating the elastic energy contribution in Eq. (9). The energy contours are calculated by changing the values of $P_1$ and $P_2$, with all the other order parameter components ($P_3$, $\theta_1$, $\theta_2$, and $\theta_3$) optimized to minimize the total free energy. As shown in Fig. 3(a), the four energy minima in the middle represent the stable rhombohedral phase, and the eight minima near the boundaries denote the metastable T-like phase. Therefore, the eighth order polynomial in Eq. (9) can describe both the rhombohedral and T-like phases [27].

To illustrate the effect of an epitaxial strain, we apply the thin film boundary conditions to the BFO system, i.e. [18,20]

$$\varepsilon_{11}=\varepsilon_{22}=\varepsilon_s, \varepsilon_{12}=0, \frac{\partial f_{BFO}}{\partial \varepsilon_{13}}=0, \frac{\partial f_{BFO}}{\partial \varepsilon_{23}}=0, \frac{\partial f_{BFO}}{\partial \varepsilon_{33}}=0, \tag{10}$$

where $\varepsilon_s$ is the biaxial epitaxial strain. From Eq. (10), three strain components are fixed, resulting in energy surfaces of different domains in a 4-D space. However, the main difference of the T and R phases is the in-plane normal strain, and to simplify the analysis, the free energy and order parameters are calculated as a function of the isotropic biaxial strain in Figs. 3(b) and 3(c).

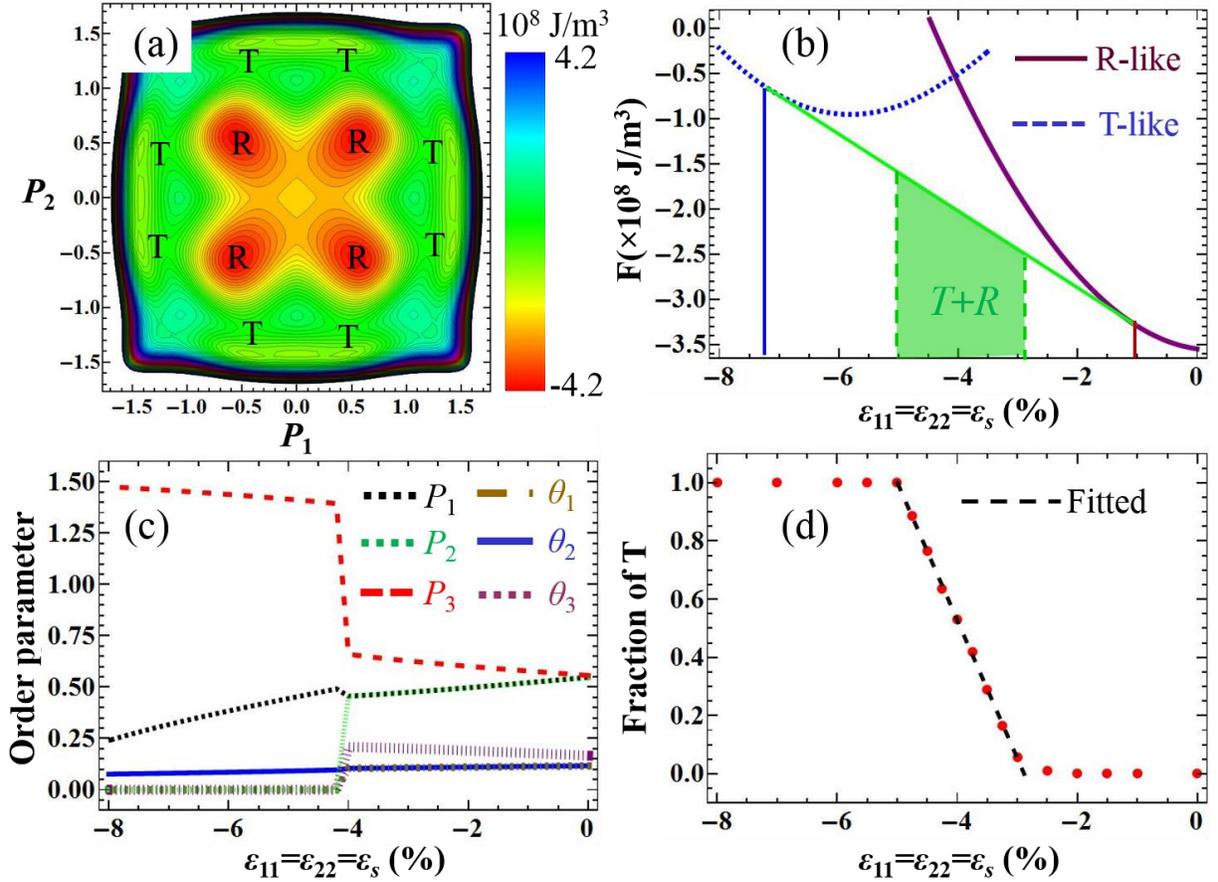

FIG. 3. Phase stability analysis of BiFeO$_3$. (a) Energy contours under stress-free boundary conditions. (b) Total free energy of the R-like and T-like phases under thin film boundary conditions. Common tangent of the two curves is plotted, and the strains of the tangent points are -1.0% and -7.2%, respectively. The shaded region represents the results from the phase-field simulations. (c) Order parameter evolution as a function of strain. (d) Volume fraction of the T-like phases from the phase-field simulations. The order parameters in (a) and (c) have SI units, i.e. *C/m²* and *rad* for $P_i$ and $\theta_i$, respectively.

The R-like phase is stable with a small strain, whereas the T-like phase is stabilized under a large compressive strain, as shown in Fig. 3(b). This is because the T-like phase has smaller in-plane lattice parameters than the R-like phases [23]. The components of the order parameters are calculated with different substrate strains. As plotted in Fig. 3(c) (note the negative sign of strain), for a compressive strain smaller than 4%, the out-of-plane component of polarization ($P_3$) increases, and the in-plane components ($P_1$ and $P_2$) decrease with an increasing strain magnitude. The tilt order parameters have similar behaviors with an increasing $\theta_3$ and decreasing $\theta_1$ and $\theta_2$. In this strain range, $P_3$ is less than 1 $C/m^2$, and $P_1$ and $P_2$ are equal, representing the R-like $M_A$ phase. With the strain larger than 4%, there is a jump of $P_3$, from 0.7 to 1.4 C/m², caused by the transition from the R-like to T-like phases. After the transition, an in-plane polarization component vanishes, and the other one has a small jump. The tilt order parameters are suppressed by the enhancement of the total polarization and only one in-plane component survives. This is consistent with first-principles calculations, which show that the T-like phase has a tilt pattern of $a^-b^0c^0$ based on the Glazer's notation [28].

To further analyze the microstructures, the coexistence of the two phases in the BFO films is predicted by the phase-field method [17]. The gradient energy and electrostatic energy are added to the total free energy, which is given by

$$F_{total} = \int [f_{BFO} + \frac{1}{2} g_{ijkl} \frac{\partial P_i}{\partial x_j} \frac{\partial P_k}{\partial x_l} + \frac{1}{2} \kappa_{ijkl} \frac{\partial \theta_i}{\partial x_j} \frac{\partial \theta_k}{\partial x_l} - E_i P_i - \frac{1}{2} \varepsilon_0 \kappa_b E_i E_i ] dV. \tag{11}$$

where $g_{ijkl}$ and $\kappa_{ijkl}$ are the gradient energy coefficients of polarization and oxygen octahedral tilt, $E_i$ is the electric field calculated by $E_i = -\frac{\partial \varphi}{\partial x_i}$ with $\varphi$ the electrostatic potential, $\varepsilon_0$ is the permittivity of free space, $\kappa_b$ is the background dielectric constant [29], and $V$ is the system volume. All the coefficients are listed in supplementary materials. Periodic boundary conditions are assumed in the $x_1$ and $x_2$ directions, and a superposition method is used for the $x_3$ direction [4]. Short-circuit electrical boundary conditions are assumed at the top and bottom surfaces. The top surface is stress-free while the bottom interface is coherently clamped by the substrate [4]. To reproduce the self-poled effect [30], $P_3$ is initially assigned with small positive random numbers, whereas the other order parameter components evolve from small random numbers. The details about how to solve the elastic equilibrium equations and Poisson equations under the above boundary conditions can be found in Ref. [4,31]. The system size is $256\Delta x \times 256\Delta x \times 20\Delta x$, and the grid spacing is $\Delta x = 0.38\ nm$.

Figure 3(d) illustrates the volume fraction change from the phase-field simulation results. The T-like phase fraction decreases linearly with respect to the strain in the mixed-phase region, i.e. following the lever rule [32]. The mixed-phase strain range in Fig. 3(d) is smaller than that from the common tangent construction in Fig. 3(b). This is because the interfacial energy and coherency strain energy were ignored in the thermodynamic analysis. Note that the lever rule is still valid after considering the contribution from coherency strain energy (see supplementary materials for details) [33].

The physical origin of the R/T two-phase mixture is similar to the formation of *a/c* domains in PTO films, i.e. to minimize the elastic energy as indicated by the common tangent construction. However, the mechanical compatibility of the domain walls is different in the two cases. The *a* and *c* domains in PTO are symmetry-related by a rotation operation, and the domain wall orientations can be obtained from either the mechanical compatibility equations or Khachatturyan's microelasticity theory, with the latter implemented in the phase-field simulations [34-36]. On the other hand, the R-like and T-like phases in BFO have distinct eigen-strains, and the domain walls are mechanically incompatible. In this case, the coherency strain energy is unavoidable, and the domain wall orientations can only be determined from the phase-field simulations by minimizing the strain energy.

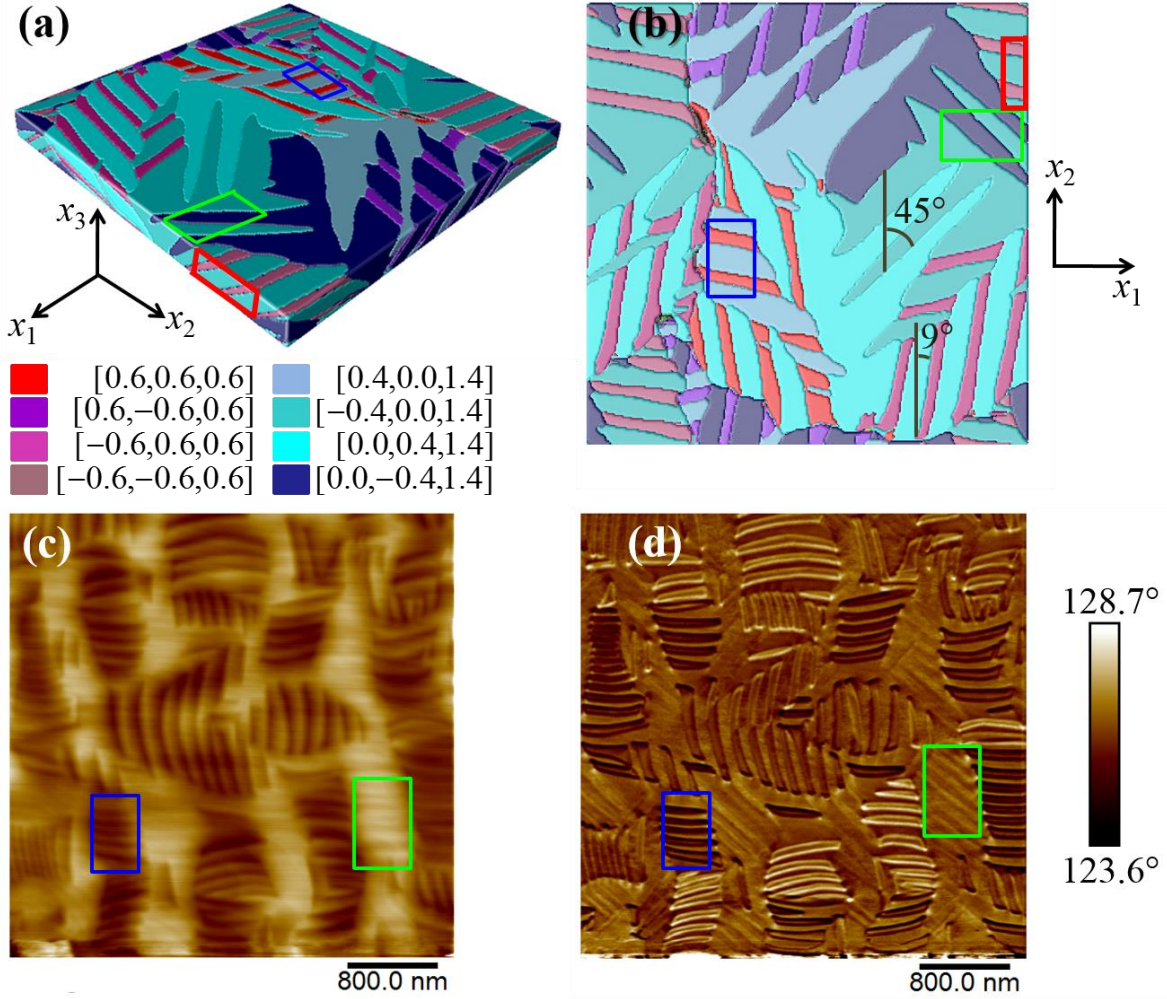

FIG. 4. Domain structures from (a) and (b) the phase-field simulations and (c) and (d) PFM measurements. (a) Three-dimensional domain structures under a compressive strain of -4.0%. (b) Corresponding domain structures in the $x_1x_2$ plane. The colors are assigned based on the polarization directions. (c) Topography of a mixed-phase BFO film with thickness of ~130 nm. (d) Corresponding in-plane PFM phase image.

The domain structures from the phase-field simulations are plotted in Figs. 4(a) and 4(b). Note that in Figs. 4(a) and 4(b), only the polarization order parameters are maintained, i.e. maintaining $\theta_i = 0 (i=1,3)$ in the simulations (see supplementary materials for the reasons). Since the initial value of $P_3$ is positive and short-circuit boundary conditions are employed, the final domains only show four R-like phase variants and four T-like variants.

A T-like domain can form two types of R/T mixed structures with the four R-like domains, e.g. a [0.4,0.0,1.4] domain forms 44° domain walls with [0.6, ±0.6, 0.6], and 67° domain walls with [-0.6, ±0.6, 0.6], with the angles measuring the polarization vector changes across the domain walls. Figures 4(a) and 4(b) show that 44° domain walls are dominant as denoted by blue rectangles, consistent with the experimental observations [37]. This demonstrates that 44° domain walls with smaller polarization change have smaller interfacial energy for the R/T mixed structures.

The determined domain wall orientations show good agreement with experimental measurements. Derived directly from the Khachatturyan's microelasticity theory (see supplementary materials for details), the domain wall normal between the [0.4, 0.0, 1.4] and [0.6, 0.6, 0.6] domains is along the [0.09, 0.58, 0.81] direction. In the $x_2x_3$ plane, the wall forms an angle of ~36° with the $x_2$ axis, close to the experimental value of ~37° [38]. In the $x_1x_2$ plane as shown in Fig. 4(b), the domain wall deviates from the $x_1$ axis by ~9°, close to the PFM measurements of ~10° in Figs. 4(c) and 4(d) [8,37,38]. The domain wall orientations of other R/T mixed structures can be obtained by symmetry operations. Also, in both the phase-field simulations and PFM measurements, the domain structures show T/T twins as denoted by green rectangles in Fig. 4.

In summary, we propose a domain/phase de-strain model for domain/phase separation in a solid under an externally imposed strain. In particular, we show that different ferroelastic domains can be treated as different phases, with a fixed overall strain component analogous to the chemical composition. The proposed model can be employed to obtain strain-temperature domain/phase diagrams and properties. This is demonstrated by the strain-temperature and strain-strain domain diagrams of $PbTiO_3$ and by the domain structures of the co-existent rhombohedral-like and tetragonal-like phases in BiFeO$_3$ films. The determined domain morphology of BiFeO$_3$ from phase-field simulations shows excellent agreement with experiments. The strain-induced stability of mixed phases is a common phenomenon in ferroelastic systems, and the proposed de-strain model provides a rather elegant and simple tool to predict the coexistence and maximum possible number of domains/phases and to design domain structures such as the types of domains/phases and their volume fractions and thus properties.

**Acknowledgements**

The work at Penn State is supported by the U.S. Department of Energy, Office of Basic Energy Sciences, Division of Materials Sciences and Engineering under Award FG02-07ER46417 (FX and LQC) and by the NSF MRSEC under Grant No. DMR-1420620 (FX and YJG). The work at Beijing Normal University is supported by the NSFC under contract numbers 11274045 and the National Basic Research Program of China under contract No.2014CB920902. This research used resources of the National Energy Research Scientific Computing Center, a DOE Office of Science User Facility supported by the Office of Science of the U.S. Department of Energy under Contract No. DE-AC02-05CH11231.

Manuscript to be summited to PRL